\documentclass[a4paper,12pt]{article}
\usepackage{epsfig}
\usepackage[dvips,usenames]{color}
\usepackage{graphicx}

\newlength{\dinwidth}
\newlength{\dinmargin}
\begin{document}
\title{Higgs boson pair production process $e^+e^-\rightarrow ZHH$ in the littlest Higgs
model at the ILC}
\bigskip
\author{Yaobei Liu$^{a}$, Linlin Du$^{b}$, Xuelei Wang$^{b}$ \\
 {\small a: Henan Institute of Science and Technology, Xinxiang
453003, P.R.China}
\thanks{E-mail:hnxxlyb2000@sina.com}\\
 {\small b: College of Physics and Information
Engineering,}\\
\small{Henan Normal University, Xinxiang  453007, P.R.China}
\thanks{This work is supported in part by the National
Natural Science
Foundation of China(Grant No.10375017 and 10575029) and a grant from Henan Institute
of Science and Technology(06040) .}\\
 }
\maketitle
\begin{abstract}
\indent   In this paper, we calculate the contribution of the
littlest Higgs(LH) model to the process $e^{+}e^{-}\rightarrow
ZHH$ at the future high energy $e^{+}e^{-}$ collider($ILC$). The
results show that, within the parameter spaces preferred by the
electroweak precision, the deviation of the total cross sections
from its $SM$ value varies from a few percent to tens percent. The
correction of the LH model to the process might be detected at the
future $ILC$ experiments in the favorable parameter space. On the
other hand, we find that the correction of the LH model is
sensitive to the trilinear Higgs coupling in some case and the
process can also provide us a chance to probe such coupling in the
LH model.
\end{abstract}
PACS number(s): 12.60Cn, 14.80.Mz, 12.15.Lk, 14.80.Cp
\newpage
\section{Introduction}~~\\
\indent The standard model(SM)provides an excellent effective field
theory description of almost all particle physics experiments. But
in the SM the Higgs boson mass suffers from an instability under
radiative corrections. The naturalness argument suggests that the
cutoff scale of the SM is not much above the electroweak scale: New
physics will appear around TeV energies. Among the extended models
beyond the SM, the little Higgs model offers a very promising
solution to the hierarchy problem in which the Higgs boson is
naturally light as a result of nonlinearly realized symmetry
\cite{little-1,little-2,little-3}. The key feature of this kind of
models is that the Higgs boson is a pseudo-Goldstone boson of an
approximate global symmetry which is spontaneously broken by a
vacuum expectation value(VEV)
at a scale of a few TeV and thus is naturally light.\\
 \indent  The most economical little Higgs model is the littlest Higgs(LH) model, which is based on the $SU(5)/SO(5)$
 nonlinear sigma model \cite{littlest}. It consists of a $SU(5)$ global
 symmetry, which is spontaneously broken down to $SO(5)$ by a vacuum
 condensate $f$. In the LH model, a set of new heavy gauge bosons$(B_{H},Z_{H},W_{H})$ and
 a new heavy-vector-like quark(T) are introduced which just cancel
 the quadratic divergence induced by SM gauge boson loops and the
 top quark loop, respectively. The distinguishing features of this
 model are the existence of these new particles and their
 couplings to the light Higgs. Measurement of these couplings would
 verify the structure of the cancelation of the Higgs mass
 quadratic divergence and prove the existence of the little Higgs
 mechanism\cite{ZHH-2}.\\
 \indent  The hunt for the Higgs boson and investigation of its properties is one of the most
  important goals of present and future
 high energy collider experiments. The precision electroweak measurement data and direct
 searches suggest that the Higgs boson must be relative light and its mass should be roughly
 in the range of 114.4GeV-208GeV at $95\%$ CL \cite{Higgs}. Studying the properties of the Higgs potential
 will reveal details of the mass-generation mechanism in spontaneously broken
 gauge theories, which can be obtained through measuring the Higgs boson self-interactions.
Recently, the Higgs boson pair production processes have been
widely considered, and the cross sections for these processes in
the SM have been evaluated at linear colliders and hadron
colliders. The phenomenology calculation show that it would be
extremely difficult to measure the Higgs self-coupling
$\lambda_{HHH}$ at the LHC \cite{LHC}, and $e^{+}e^{-}$ linear
colliders, where the study of the $e^{+}e^{-}\rightarrow ZHH$ and
$HH\nu\overline{\nu}$ can be performed with good accuracy,
represent a possibly unique opportunity for performing the study
of the trilinear Higgs self-coupling \cite{ZHH-1,ZHH3,self}. For
the center of mass(c.m.)energy $\sqrt{s}$ from 500 GeV up to 1
TeV, the $ZHH$ production with intermediate Higgs boson mass is
the most promising process among the various Higgs
doublet-production processes. Since the cross section is
relatively large and all the final states can be identified
without large missing momentum, the process $e^{+}e^{-}\rightarrow
ZHH$ is the best one among the various Higgs doublet-production
processes to look for the Higgs self-coupling
during the first stage of the future linear collider.\\
 \indent  We know that the most important Higgs production process at the linear collider
 is the Higgs-strahlung process $e^{+}e^{-}\rightarrow ZH$.
 The correction effects of LH model to this process was studied in Ref.\cite{115004}. It is found that the
 correction effects mainly come from the heavy gauge boson $B_{H}$, in most
 parameter space, the deviation of the total cross section from its SM value is larger
 than $5\%$, which may be detected at the future ILC experiment. However, the
double Higgs-strahlung process $e^{+}e^{-}\rightarrow ZHH$
includes the trilinear Higgs coupling which is different from the
process $e^{+}e^{-}\rightarrow ZH$. For the process
$e^{+}e^{-}\rightarrow ZHH$, the
 contribution of the LH model comes from not only the new heavy
 gauge bosons $B_{H},Z_{H}$ but also the modification of the self-couplings of Higgs
 boson. So the process $e^{+}e^{-}\rightarrow ZHH$ can also provide
 some useful information about the modification of trilinear Higgs coupling in
 the LH model to complement the study of the process $e^{+}e^{-}\rightarrow
 ZH$.
 In this paper, we
 consider the double Higgs-strahlung process $e^{+}e^{-}\rightarrow ZHH$ and study whether
 the correction effects of LH model to this process can be detected at the future ILC
 experiment.

This paper is organized as follows, In section two, we first
briefly introduce the LH model, and then give the production
amplitude of the process. The numerical results and discussions
are presented in section three. Our conclusions are given in
section four.
\section{The process $e^{+}e^{-}\rightarrow ZHH$
in the $LH$ model}~~\\
\indent The LH model is based on the $SU(5)/SO(5)$ nonlinear sigma
model. At the scale $\Lambda_{s}\sim4\pi$$f$, the global $SU(5)$
symmetry is broken into its subgroup $SO(5)$ via a vacuum
condensate $f$, resulting in 14 Goldstone bosons. The effective
field theory of these Goldstone bosons is parameterized by a
non-linear $\sigma$ model with gauged symmetry $[SU(2)\times
U(1)]^{2}$, spontaneously broken down to its diagonal subgroup
$SU(2)\times U(1)$ which is identified as the SM electroweak gauge
group. Four of these Goldstone bosons are eaten by the broken
gauge generators, leaving 10 states that transform under the SM
gauge group as a doublet H and a triplet $\Phi$. This breaking
scenario also gives rise to four massive gauge bosons $B_{H}$,
$Z_{H}$ and $W^{\pm}_{H}$, which might produce characteristic
signatures in the present and future high energy
collider experiments \cite{signatures-1,signatures-2,signatures-3}.\\
 \indent  After EWSB, the final mass eigenstates are obtained via the
 mixing between the heavy and light gauge bosons.
 They include the light (SM-like) bosons $Z_{L}$,$A_{L}$ and
$W^{\pm}_{L}$ observed at experiments, and new heavy bosons
$Z_{H}$,$B_{H}$ and $W^{\pm}_{H}$ that could be observed at future
experiments. The masses of neutral gauge bosons are given to ${\cal
O}(v^{2}/f^{2})$ by \cite{signatures-4}
\begin{eqnarray}
M^{2}_{A_{L}}&=&0,\\
M^{2}_{Z_{L}}&=&(M^{SM}_{Z})^{2}\{1-\frac{v^{2}}{f^{2}}[\frac{1}{6}+\frac{1}{4}(c^{2}-s^{2})^{2}+
\frac{5}{4}(c'^{2}-s'^{2})^{2}-\frac{x^{2}}{2}]\},\\
M^{2}_{Z_{H}}&=&(M^{SM}_{Z})^{2}c^{2}_{W}\{\frac{f^{2}}{s^{2}c^{2}v^{2}}-1+\frac{v^{2}}{2f^{2}}[
\frac{(c^{2}-s^{2})^{2}}{2c^{2}_{W}}+\chi_{H}\frac{g'}{g}\frac{c'^{2}s^{2}+c^{2}s'^{2}}{cc'ss'}]\},\\
M^{2}_{B_{H}}&=&(M^{SM}_{Z})^{2}s^{2}_{W}\{\frac{f^{2}}{5s'^{2}c'^{2}v^{2}}-1+\frac{v^{2}}{2f^{2}}[
\frac{5(c'^{2}-s'^{2})^{2}}{2s^{2}_{W}}-\chi_{H}\frac{g'}{g}\frac{c'^{2}s^{2}+c^{2}s'^{2}}{cc'ss'}]\},
\end{eqnarray}
with $x=\frac{4fv'}{v^{2}},~
\chi_{H}=\frac{5}{2}gg'\frac{scs'c'(c^{2}s'^{2}+s^{2}c'^{2})}{5g^{2}s'^{2}c'^{2}-g's^{2}c^{2}}$,
where~$v$=246 GeV is the elecroweak scale, $v'$ is the VEV of the
scalar $SU(2)_{L}$ triplet and $s_{W}(c_{W})$ represents the
sine(cosine) of the weak mixing angle.
 $c(s=\sqrt{1-c^{2}})$ is the mixing parameter between $SU(2)_{1}$ and $SU(2)_{2}$
 gauge bosons and the mixing parameter $c'(s'=\sqrt{1-c'^{2}})$ comes
 from the mixing between $U(1)_{1}$ and $U(1)_{2}$ gauge bosons. Using these mixing parameters, we
can represent the SM gauge coupling constants as $g=g_{1}s=g_{2}c$
and $g'=g_{1}s'=g_{2}c'$. The mass of neutral scalar boson
$M_{\Phi^{0}}$ can be given as \cite{signatures-1}
\begin{eqnarray}
M^{2}_{\Phi^{0}}=\frac{2m^{2}_{H^{0}}f^{2}}{v^{2}(1-x^{2})}.
\end{eqnarray}
The above equation about the mass of $\Phi$ requires a constraint of
0$\leq$x$<$1, which shows the relation
between the scale $f$ and the VEV of the Higgs field doublet and triplet$(v,v')$.\\
\indent  Taking account of the gauge invariance of the Yukawa
couplings and the $U(1)$ anomaly cancellation, one can write the
couplings of the neutral gauge bosons $V_i(V_i=Z_{L},B_{H},Z_{H})$
to electrons pair in the form of
$\wedge_{\mu}^{V_i\bar{e}e}=i\gamma_{\mu}(g^{V_i\bar{e}e}_{V}+g^{V_i\bar{e}e}_{A}\gamma^{5})$
with \cite{signatures-1}
\begin{eqnarray}
g^{Z_{L}\bar{e}e}_{V}&=&-\frac{e}{4s_{W}c_{W}}\{(-1+4s^{2}_{W})-\frac{v^{2}}
{f^{2}}[\frac{1}{2}c^{2}(c^{2}-s^{2})-\frac{15}{2}(c'^{2}-s'^{2})(c'^{2}-\frac{2}{5})]\},\\
\nonumber
g^{Z_{L}\bar{e}e}_{A}&=&-\frac{e}{4s_{W}c_{W}}\{1+\frac{v^{2}}{f^{2}}[\frac{1}{2}c^{2}(c^{2}-s^{2})
+\frac{5}{2}(c'^{2}-s'^{2})(c'^{2}-\frac{2}{5})]\},\\ \nonumber
g^{Z_{H}\bar{e}e}_{V}&=&-\frac{ec}{4s_{W}s},
 \hspace{5cm}g^{Z_{H}\bar{e}e}_{A}=\frac{ec}{4s_{W}s},\\ \nonumber
g^{B_{H}\bar{e}e}_{V}&=&\frac{e}{2c_{W}s'c'}(\frac{3}{2}c'^{2}-\frac{3}{5}),
 \hspace{3cm}g^{B_{H}\bar{e}e}_{A}=\frac{e}{2c_{W}s'c'}(\frac{1}{2}c'^{2}-\frac{1}{5}).
\end{eqnarray}
The couplings of the gauge bosons to Higgs boson and self-Higgs
coupling can be written as
\begin{eqnarray}
g^{Z_{L\mu}Z_{L\nu}H}&=&\frac{ie^{2}vg_{\mu\nu}}{2s^{2}_{W}c^{2}_{W}}\{1-\frac{v^{2}}
{f^{2}}[\frac{1}{3}-\frac{3}{4}x^{2}+\frac{1}{2}(c^{2}-s^{2})^{2}+\frac{5}
{2}(c'^{2}-s'^{2})^{2}]\},\\ \nonumber
g^{Z_{H\mu}Z_{H\nu}H}&=&-\frac{ie^{2}}{2s_{W}^{2}}vg_{\mu\nu},
\hspace{4.0cm}g^{B_{H\mu}B_{H\nu}H}=-\frac{ie^{2}}{2c_{W}^{2}}vg_{\mu\nu},\\
\nonumber
g^{Z_{L\mu}Z_{H\nu}H}&=&-\frac{ie^{2}(c^{2}-s^{2})vg_{\mu\nu}}{4s^{2}_{W}c_{W}sc},
 \hspace{2.7cm}g^{Z_{L\mu}B_{H\nu}H}=-\frac{ie^{2}(c'^{2}-s'^{2})vg_{\mu\nu}}{4s_{W}c^{2}_{W}s'c'},\\
 \nonumber
g^{Z_{H\mu}B_{H\nu}H}&=&-\frac{ie^{2}vg_{\mu\nu}}{4s_{W}c_{W}}\frac{(c^{2}s'^{2}+s^{2}c'^{2})}{scs'c'},
 \hspace{1.5cm}g^{Z_{L\mu}Z_{L\nu}HH}=\frac{ie^{2}vg_{\mu\nu}}{2s^{2}_{W}c^{2}_{W}},\\ \nonumber
g^{Z_{L\mu}Z_{H\nu}HH}&=&-\frac{ie^{2}(c^{2}-s^{2})g_{\mu\nu}}{4s^{2}_{W}c_{W}sc},
 \hspace{2.7cm}g^{Z_{L\mu}B_{H\nu}HH}=-\frac{ie^{2}(c'^{2}-s'^{2})g_{\mu\nu}}{4s_{W}c^{2}_{W}s'c'},\\ \nonumber
g^{HHH}&=&-\frac{i3m_{H}^{2}}{v}[1-\frac{11v^{2}x^{2}}{4f^{2}(1-x^{2})}].
\end{eqnarray}
 \indent  In the LH model, the heavy triple Higgs boson $\Phi^{0}$ exchange
can also contribute to the process $e^{+}e^{-}\rightarrow ZHH$.
However, compared to the contributions coming from the new gauge
bosons, the contribution of $\Phi^{0}$ exchanging is suppressed by
the order $v^{4}/f^{4}$, which can be seen from the couplings
between gauge bosons and scalars\cite{signatures-1}. Thus, we can
ignore the contribution of the scalar triplets to the process
$e^{+}e^{-}\rightarrow ZHH$.

 The relevant Feynman diagrams for the process $e^{+}e^{-}\rightarrow ZHH$ in the LH model are
 shown in Fig.1 at the tree-level.
\begin{figure}[h]
\begin{center}
\epsfig{file=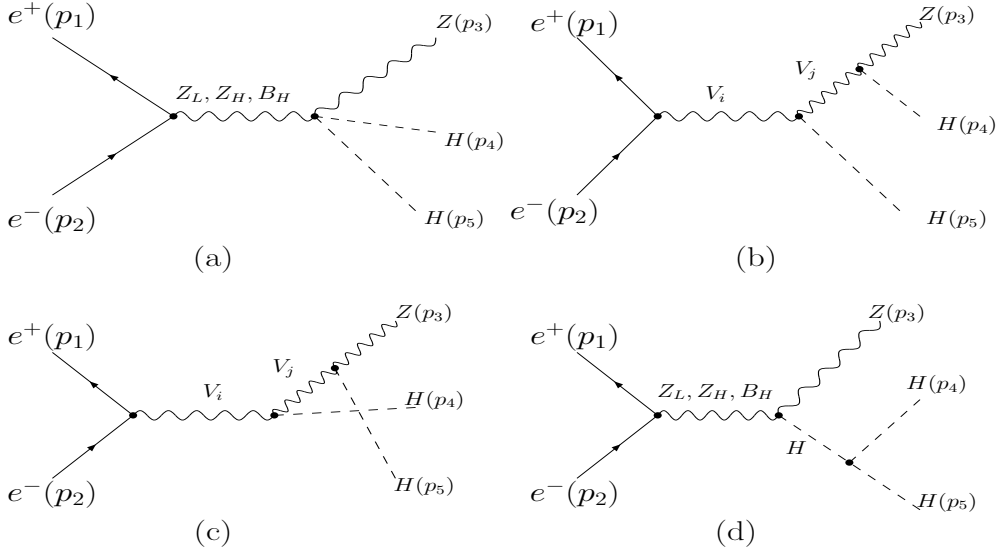,width=450pt,height=500pt} \vspace{-8cm}
\caption{\small The Feynman diagrams of the process
 $e^{+}e^{-}\rightarrow ZHH$ in the $LH$ model.} \label{fig1}
\end{center}
\end{figure}

 The invariant production amplitudes of the process
 can be written as
\begin{equation}
 M=\sum_{V_{i}=Z_{L},Z_{H},B_{H}}M_{a}^{V_{i}}+
 \sum_{V_{i,j}=Z_{L},Z_{H},B_{H}}M_{b}^{V_{i}V_{j}}+
 \sum_{V_{i,j}=Z_{L},Z_{H},B_{H}}M_{c}^{V_{i}V_{j}}+
 \sum_{V_{i}=Z_{L},Z_{H},B_{H}}M_{d}^{V_{i}},
 \end{equation}
 with
 \begin{eqnarray}
 M_{a}^{V_{i}}&=&\overline{v_{e}}(p_{1})\wedge_{\mu}^{V_{i}e\overline{e}}u_{e}(p_{2})
 G^{\mu\nu}(p_{1}+p_{2}, M_{V_{i}})\wedge_{\nu\rho}^{HHZV_{i}}\varepsilon^{\rho}(p_{3}),\\
 \nonumber
 M_{b}^{V_{i}V_{j}}&=&\overline{v_{e}}(p_{1})\wedge_{\mu}^{V_{i}e\overline{e}}u_{e}(p_{2})
 G^{\mu\nu}(p_{1}+p_{2}, M_{V_{i}})\wedge_{\nu\rho}^{HV_{i}V_{j}}G^{\rho\lambda}(p_{3}+p_{4}, M_{V_{j}})
 \wedge_{\lambda\tau}^{HZV_{j}}\varepsilon^{\tau}(p_{3}),\\ \nonumber
 M_{c}^{V_{i}V_{j}}&=&\overline{v_{e}}(p_{1})\wedge_{\mu}^{V_{i}e\overline{e}}u_{e}(p_{2})
 G^{\mu\nu}(p_{1}+p_{2}, M_{V_{i}})\wedge_{\nu\rho}^{HV_{i}V_{j}}G^{\rho\lambda}(p_{3}+p_{5}, M_{V_{j}})
 \wedge_{\lambda\tau}^{HZV_{j}}\varepsilon^{\tau}(p_{3}),\\ \nonumber
 M_{d}^{V_{i}}&=&-\overline{v_{e}}(p_{1})\wedge_{\mu}^{V_{i}e\overline{e}}u_{e}(p_{2})
 G^{\mu\nu}(p_{1}+p_{2}, M_{V_{i}})\wedge_{\nu\rho}^{HZV_{i}}G(p_{4}+p_{5},
 M_{H})
 g^{HHH}\varepsilon^{\rho}(p_{3}).
 \end{eqnarray}
 Here, $G^{\mu\nu}(p,M)=\frac{-ig^{\mu\nu}}{p^{2}-M^{2}}$ is the propagator of the particle.
 We can see that this process in the LH model receives additional
 contributions from the heavy gauge bosons $Z_H,B_H$. Furthermore,
 the modification of the relations among the SM parameters and the
 precision electroweak input parameters, the correction terms to the
 SM $HHH$ coupling can also produce corrections to this process. In our numerical calculation, we will
also take into account these correction effects. The main decay
modes of $B_H$ and $Z_H$ are $V_i\rightarrow f\bar{f}$($f$
represents any quarks and leptons in the SM) and $V_i\rightarrow
ZH$. The decay widths of these modes have been explicitly given in
references
\cite{signatures-1,yue}.\\
\indent With above production amplitudes, we can obtain the
production cross section directly. In the calculation of the cross
section, instead of calculating the square of the amplitudes
analytically, we calculate the amplitudes numerically by using the
method of the references\cite{HZ} which can greatly simplify our
calculation.
\section{ The numerical results and discussions}\
\indent In the numerical calculation, we take the input parameters
as
 $M_{Z}^{SM}=91.187$ GeV, $s_{W}^{2}=0.2315$ \cite{data}.
 For the light Higgs boson H, in this paper,
we only take the illustrative value $M_{H}=120$ GeV. In this case,
the possible decay modes of H are $b\bar{b}$, $c\bar{c}$,
$l\bar{l}$[l=$\tau$, $\mu$ or e], $gg$ and $\gamma\gamma$. However,
the total decay width $\Gamma_{H}$ is dominated by the decay channel
$H\rightarrow b\bar{b}$. In the LH model, $\Gamma_{H}$ is modified
from that in the SM by the order of $v^{2}/f^{2}$ and has been
studied in Ref.\cite{bb}. The c.m. energy
of the ILC is assumed as $\sqrt{s}$=500 GeV.\\
\indent The absence of custodial $SU(2)$ global symmetry in the LH
model yields weak isospin violating contributions  to the
electroweak precision observables. In the early study, global fits
to the experimental data put rather severe constraints in the
$f>4$ TeV at $95\%$ C.L.\cite{cons1,con1}. However, their analysis
is based on a simple assumption that the SM fermions are charged
only under $U(1)_{1}$. If the SM fermions are charged under
$U(1)_{1}\times U(1)_{2}$, the constraints become relaxed: the
substantial parameter space allows $f=1\sim2$ TeV \cite{cons2}. If
only the $U(1)_{Y}$ is gauged, the experimental constraints are
looser \cite{cons2,cons3}. Therefore , the new contributions are
suppressed: $f=1\sim2$ TeV allowed for the mixing
 parameters $c$ and $c^{'}$ in the ranges of
 $0 \sim 0.5,0.62 \sim 0.73$ \cite{cons2,constraints}.
 The parameter $x<1$ parameterizes the ratio of the triplet and doublet
 VEV's. Taking into account the constraints on $f,c,c',x$, we
 take them as the free parameters in our numerical calculation.
 The numerical results are summarized in Figs.(2-4).
\begin{figure}[h]
\begin{center}
\scalebox{0.85}{\epsfig{file=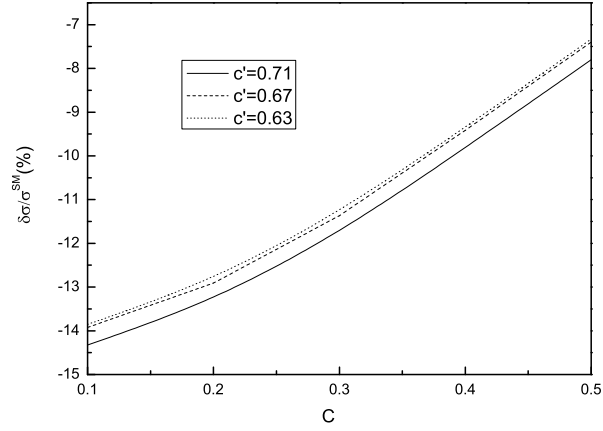}}\\
\end{center}
\caption{\small The relative correction $\delta\sigma/\sigma^{SM}$
as a function of the mixing parameter c for f=1 TeV, $x$=0.5,
$M_{H}=120$ GeV and different values of the mixing parameter $c'$.}
\end{figure}

\indent  The relative correction $\delta\sigma/\sigma^{SM}$ is
plotted in Fig.2 as a function of the mixing parameter c for $f$=1
TeV, $x=0.5$, $M_{H}=120$ GeV, $c'=0.63,0.67,0.71$, respectively.
In Fig.2, $\delta\sigma=\sigma^{tot}-\sigma^{SM}$ and
$\sigma^{SM}$ is the tree-level cross section of the $ZHH$
production predicted by the SM.
 From Fig.2, we can see that the absolute value of the relative correction
 decreases with the mixing parameter c increasing. For
 $x=0.5$,
 the absolute value of $\delta\sigma/\sigma^{SM}$ is in the range of
 $8\%-14\%$ in the most
parameter space limited by the electroweak precision data. The
curves also show that with an increase of the value of $c'$,
 the effect of the LH
 model is getting stronger. For $f<3$ TeV, the mass of $B_{H}$ may be lighter than
500 GeV\cite{signatures-3}. In most parameter spaces of the LH
model, the mass of the heavy gauge boson $Z_{H}$ is larger than 1
TeV. So, there is no s-channel resonance effects in our numerical
results.\\
\begin{figure}[b]
\begin{center}
\scalebox{0.85}{\epsfig{file=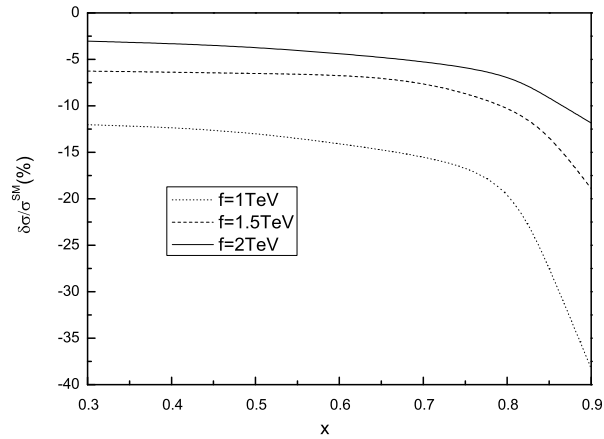}}
\end{center}
\caption{\small The relative correction $\delta\sigma/\sigma^{SM}$
as a function of the mixing parameter $x$ for c=0.3, $c'$=0.67,
and $f=1,1.5,2$ TeV, respectively.}
\end{figure}
\indent To see the dependence of the relative correction on the
parameter $x$, in Fig.3, we plot $\delta\sigma/\sigma^{SM}$ as a
function of the mixing parameter $x$ for c=0.3, $c'=0.67$ and
three values of the scale parameter $f$. From Fig.3 we can see
that, the absolute
 value of the relative correction decreases as $f$ increasing. The curves also demonstrate that
the effect of the LH model is not sensitive to $x$ in the range of
$x\leq 0.75$. This is because the deviations of the cross section
from the SM are mainly aroused by the contributions of the new gauge
bosons when $x\leq 0.75$. However, the figure shows that the
absolute values of $\delta\sigma/\sigma^{SM}$ raised quickly when we
take the $x\rightarrow 1$ limit and in this case the main
contribution to the cross section comes from the Feynman diagram
involving the trilinear interaction of the SM Higgs boson, which is
consistent with the conclusions for the contributions of the LH
model to Higgs boson pair production at hadron colliders
\cite{hadron}. So, if $x$ is large enough the significant correction
of the LH model to the trilinear Higgs coupling should be
observable. The process $e^+e^-\rightarrow ZHH$ can open a unique
window to probe the Higgs self-coupling which can complement the
process $e^+e^-\rightarrow ZH$.

\begin{figure}[hb]
\begin{tabular}{cc}
~~~~~~~~~~~~~~~~~~~~~~~\scalebox{0.85}{\epsfig{file=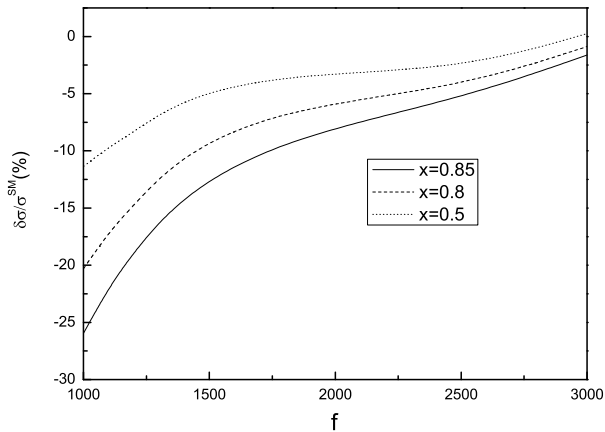}}
\end{tabular}
\caption{\small The relative correction $\delta\sigma/\sigma^{SM}$
as a function of the the scale parameter $f$ for c=0.3, $c'=0.67$,
$M_{H}=120$ GeV,
 and three values of the mixing parameter $x$.}
\end{figure}

In general, the contributions of the LH model to the observables
are dependent on the factor $1/f^{2}$. In order to obtain the
generic conclusion, we also plot $\delta\sigma/\sigma^{SM}$ as a
function of $f$(1-3 TeV) for three values of the parameter $x$ and
take $c=0.3$, $c^{'}=0.68$ in Fig 4. One can see that the absolute
relative correction drops sharply with $f$ increasing, which is
consistent with the conclusions for the corrections of the LH
model to other observables. On the other hand, we can see that the
absolute relative correction increases as the parameter $x$
increasing. For example, the absolute relative
correction may reach about $20\%$ when $x=0.8$ and $f=1$ TeV.\\
\indent As has been mentioned above, the total cross section of
$e^{+}e^{-}\rightarrow ZHH$ can reach the order of $10^{-1}$ fb at
the ILC. This cross section amounts to about 100 events with the
integrated luminosity of 1000 $fb^{-1}$. The $1\sigma$ statistical
error corresponds to about $10\%$ precision. The reference\cite{ILC}
have reviewed the expected experimental precision with which the ZHH
cross section can be measured. Even we consider the systemic error
of the ILC, the ILC can measure the cross section with the precision
of $17\%$ assuming a 120 GeV Higgs and the integrated luminosity
1000 $fb^{-1}$ at 500 GeV. The sensitivity can be further improved
when a multi-variable selection based on a neural network is applied
which can reduce the uncertainty from $17\%$ to $13\%$. The relative
correction of the LH model to the cross section is only comparable
to the ILC measurement precision and might be detected at the ILC in
the favorable parameter spaces(for example, small value of $f$)
preferred by the electroweak precision. The statistical acuracy to
measure the trilinear Higgs coupling is $22\%$ for $M_H=120$ GeV
with an integrated luminosity of $1000 fb^{-1}$, using the neural
network selection\cite{ILC}. Only for small $f$ and large $x$, the
correction of the LH model to the trilinear Higgs coupling can be
detected.
\section{Conclusion}
\indent The little Higgs model, which can solve the hierarchy
problem, is a promising alternative model of new physics beyond
the standard model. Among various little Higgs models, the
littlest Higgs(LH) model is one of the simplest and
phenomenologically viable models. The distinguishing feature of
this model is the existence of the new scalars, the new gauge
bosons, and the vector-like top quark. These new particles
contribute to the experimental observables which could provide
some clues of the existence of the LH model. In this paper, we
study the potential to detect the contribution of the LH model via
the process
$e^+e^-\rightarrow ZHH$ at the future $ILC$ experiments.\\
\indent In the parameter spaces($f=1\sim2$ TeV, $c=0\sim0.5$,
$c'=0.62\sim0.73$) limited by the electroweak precision data, we
calculate the cross section correction of the LH model to the
process $e^{+}e^{-}\rightarrow ZHH$. We find that the correction
is significant even when we consider the constraint of electroweak
precision data on the parameters. The relative correction varies
from a few percent to tens of percents. The LH model is a weak
interaction theory and it is hard to detect its contributions and
measure its couplings at the LHC. With the high c.m. energy and
luminosity, the future ILC will open an ideal window to probe into
the LH model and study its properties. In some favorable case, the
relative correction of the LH model to the process
$e^{+}e^{-}\rightarrow ZHH$ might be large enough to be measured
with high precision at the ILC. Furthermore, the process can also
open a unique window into the trilinear Higgs coupling in LH
model.

\end{document}